\documentclass{new_tlp}

\usepackage{url}

  \title[Theory and Practice of Logic Programming]
        {Introduction to the 35th International Conference on Logic Programming Special Issue}

\author[E.~Erdem and A.~Formisano and G.~Vidal and F.~Yang]
{ESRA ERDEM \\
  Sabanci University, Turkey\\
  \email{esraerdem@sabanciuniv.edu}
  \and
  ANDREA FORMISANO\\
  Universit\`a di Perugia, Italy\\
  \email{andrea.formisano@unipg.it}
  \and
  GERM\'AN VIDAL \\
MiST, VRAIN, Universitat Polit\`ecnica de Val\`encia, Spain\\
\email{gvidal@dsic.upv.es}
\and
FANGKAI YANG\\
NVIDIA Corporation, USA\\
\email{fangkaiy@nvidia.com}
}

\jdate{March 2003}
\pubyear{2003}
\pagerange{\pageref{firstpage}--\pageref{lastpage}}
\doi{S1471068401001193}

\begin{document}

\label{firstpage}

\maketitle
%

This volume contains the Regular Papers, Technical Communications and the Doctoral
Consortium papers of the 35th International Conference on Logic
Programming (ICLP 2019), held in Las Cruces, New Mexico, USA, from
September 20--25, 2019.

Since the first conference held in Marseille in 1982, ICLP has been
the premier international event for presenting research in logic
programming.  Contributions are sought in all areas of logic
programming, including but not restricted to:
\begin{description}
\item[Foundations:] Semantics, Formalisms, Nonmonotonic reasoning,
  Knowledge representation.
\item[Languages:] Concurrency, Objects, Coordination, Mobility, Higher
  Order, Types, Modes, Assertions, Modules, Meta-programming,
  Logic-based domain-specific languages, Programming Techniques.
\item[Declarative programming:] Declarative program development,
  Analysis, Type and mode inference, Partial evaluation, Abstract
  interpretation, Transformation, Validation, Verification, Debugging,
  Profiling, Testing, Execution visualization
\item[Implementation:] Virtual machines, Compilation, Memory
  management, Parallel or distributed execution, Constraint handling
  rules, Tabling, Foreign interfaces, User interfaces.
\item[Related Paradigms and Synergies:] Inductive and Co-inductive
  Logic Programming, Constraint Logic Programming, Answer Set
  Programming, Interaction with SAT, SMT and CSP solvers, Logic
  programming techniques for type inference and theorem proving,
  Argumentation, Probabilistic Logic Programming, Relations to
  object-oriented and Functional programming.
\item[Applications:] Databases, Big Data, Data integration and
  federation, Software engineering, Natural language processing, Web
  and Semantic Web, Agents, Artificial intelligence, Computational
  life sciences, Education, Cybersecurity, and Robotics.
\end{description}


Besides the main track, ICLP 2019 included the following additional
tracks and special sessions:
\begin{itemize}
\item \textbf{Applications Track:} This track invited submissions of
  papers on emerging and deployed applications of logic programming,
  describing all aspects of the development, deployment, and
  evaluation of logic programming systems to solve real-world
  problems, including interesting case studies and benchmarks, and
  discussing lessons learned.
	
\item \textbf{Sister Conferences and Journal Presentation Track:} This
  track provided a forum to discuss important results related to logic
  programming that appeared recently (from January 2017 onwards) in
  selective journals and conferences, but have not been previously
  presented at ICLP.

\item \textbf{Research Challenges in Logic Programming Track:} This
  track invited submissions of papers describing research challenges
  that an individual researcher or a research group is currently
  attacking. The goal of the track is to promote discussions, exchange
  of ideas, and possibly stimulate new collaborations.

\item \textbf{Special Session. Women in Logic Programming:} This
  special session included an invited talk and presentations
  describing research by women in logic programming.
\end{itemize}

The organizers of ICLP 2019 were:
\begin{itemize}
\item[] \textbf{General Chairs}
  \begin{description}
  \item[] Enrico Pontelli, New Mexico State University
  \item[] Tran Cao Son, New Mexico State University
  \end{description}
\item[] \textbf{Program Chairs}
  \begin{description}
  \item[] Esra Erdem, Sabanci University
  \item[] Germ\'an Vidal, Universitat Polit\`ecnica de Val\`encia
  \end{description}
\item[] \textbf{Publicity Chair}
  \begin{description}
  \item[] Ferdinando Fioretto, Georgia Institute of Technology
  \end{description}
\item[] \textbf{Workshops Chair}
\begin{description}
\item[] Martin Gebser, University of Klagenfurt and Graz University of
  Technology
\end{description}
\item[] \textbf{Tutorials Chair}
\begin{description}
\item[] Pedro Cabalar, University of Corunna
\end{description}
\item[] \textbf{DC Chairs}
\begin{description}
\item[] Paul Fodor, Stony Brook New York
\item[] Daniela Inclezan, Miami University
\end{description}
\item[] \textbf{Programming Competition Chairs}
\begin{description}
\item[] Jos\'e Morales, IMDEA Software Institute
  \item[] Orkunt Sabuncu, TED University
\end{description}
\item[] \textbf{Applications Track Chairs}
\begin{description}
\item[] Andrea Formisano, Universita' di Perugia
  \item[] Fangkai Yang, NVIDIA
  Corporation
\end{description}
\item[] \textbf{Sister Conferences and Journal Presentation Track Chairs}
\begin{description}
\item[] Bart Bogaerts, Vrije Universiteit Brussel
  \item[] Giovambattista Ianni, Universit\`a della
  Calabria
\end{description}
\item[] \textbf{Research Challenges in Logic Programming Track Chairs}
\begin{description}
\item[] Alessandro dal Pal\`u, Universit\`a di Parma
  \item[] Amelia Harrison,
    University of Texas at Austin and Google Inc.
    \item[] Joohyung Lee, Arizona
  State University
\end{description}
\item[] \textbf{Women in Logic Programming Special Session Chairs}
  \begin{description}
  \item[] Alicia Villanueva, Universitat Polit\`ecnica de Val\`encia
  \item[] Marina De Vos, University of Bath
  \end{description}
\end{itemize}

Three kinds of submissions were accepted:
\begin{itemize}
\item Technical papers for technically sound, innovative ideas that
  can advance the state of logic programming.
\item Application papers that impact interesting application domains.
\item System and tool papers which emphasize novelty, practicality,
  usability, and availability of the systems and tools described.
\end{itemize}

ICLP implemented the hybrid publication model used in all recent
editions of the conference, with journal papers and Technical
Communications (TCs), following a decision made in 2010 by the
Association for Logic Programming. Papers of the highest quality were
selected to be published as rapid publications in this special issue of
TPLP. The TCs comprise papers which the Program Committee (PC) judged
of good quality but not yet of the standard required to be accepted
and published in TPLP as well as extended abstracts from the different
tracks and dissertation project descriptions
stemming from the Doctoral Program (DP) held with ICLP.

We have received 122 submissions of abstracts, of which 92 resulted in
full submissions, distributed as follows: ICLP main track (59),
Applications track (9 full papers and 9 short papers), Sister
Conferences and Journal Presentation track (11), and Women in Logic
Programming session (4). The Program Chairs organized the refereeing
process, which was undertaken by the PC with the support of external
reviewers. Each technical paper was reviewed by at least three
referees who provided detailed written evaluations. This enabled a
list of papers to be short-listed as candidates for rapid
communication. The authors of these papers revised their submissions
in light of the reviewers’ suggestions, and all these papers were
subject to a second round of reviewing. Of these candidates papers, 30
were accepted as rapid communications, to appear in the special
issue.
In addition, the PC recommended 45 papers to be accepted as
technical communications, either as full papers or extended abstracts,
of which 44 were also presented at the conference (1 was withdrawn).

The 30 rapid communications that appear in this special issue are
listed below:
\begin{itemize}
\item \emph{Tiantian Gao, Paul Fodor and Michael Kifer}. Querying
  Knowledge via Multi-Hop English Questions.\\
\url{https://arxiv.org/abs/1907.08176}

\item \emph{Mario Alviano, Nicola Leone, Pierfrancesco Veltri and
    Jessica Zangari}. Enhancing magic sets with an application to
  ontological reasoning.\\
\url{https://arxiv.org/abs/1907.08424}

\item \emph{Jorge Fandinno}. Founded (Auto)Epistemic Equilibrium Logic
  Satisfies Epistemic Splitting.\\
\url{https://arxiv.org/abs/1907.09247}

\item \emph{Giovanni Amendola and Francesco Ricca}. Paracoherent
  Answer Set Semantics meets Argumentation Frameworks.\\
\url{https://arxiv.org/abs/1907.09426}

\item \emph{Giovanni Amendola, Francesco Ricca and Mirek
    Truszczynski}. Beyond NP: Quantifying over Answer Sets.\\
\url{https://arxiv.org/abs/1907.09559}

\item \emph{Elvira Albert, Miquel Bofill, Cristina Borralleras,
    Enrique Martin-Martin and Albert Rubio}. Resource Analysis driven
  by (Conditional) Termination Proofs.\\
\url{https://arxiv.org/abs/1907.10096}

\item \emph{Giovanni Amendola, Carmine Dodaro and Marco
    Maratea}. Abstract Solvers for Computing Cautious Consequences of
  ASP programs.\\
\url{https://arxiv.org/abs/1907.09402}

\item \emph{Giovanni Amendola, Carmine Dodaro and Francesco
    Ricca}. Better Paracoherent Answer Sets with Less Resources.\\
\url{https://arxiv.org/abs/1907.09560}

\item \emph{Gonzague Yernaux and Wim Vanhoof}. Anti-unification in
  Constraint Logic Programming.\\
\url{https://arxiv.org/abs/1907.10333}

\item \emph{Stefania Costantini}. About epistemic negation and world
  views in Epistemic Logic Programs.\\
\url{https://arxiv.org/abs/1907.09867}

\item \emph{Fernando S\'aenz-P\'erez}. Applying Constraint Logic
  Programming to SQL Semantic Analysis.\\
\url{https://arxiv.org/abs/1907.10914}

\item \emph{Wolfgang Faber, Michael Morak and Stefan Woltran}. On
  Uniform Equivalence of Epistemic Logic Programs.\\
\url{https://arxiv.org/abs/1907.10925}

\item \emph{Efthimis Tsilionis, Nikolaos Koutroumanis, Panagiotis
    Nikitopoulos, Christos Doulkeridis and Alexander Artikis}. Online
  Event Recognition from Moving Vehicles: Application Paper.\\
\url{https://arxiv.org/abs/1907.11007}

\item \emph{Bernardo Cuteri, Carmine Dodaro, Francesco Ricca and Peter
    Sch\"uller}. Partial Compilation of ASP Programs.\\
\url{https://arxiv.org/abs/1907.10469}

\item \emph{Mar\'{\i}a Alpuente, Demis Ballis, Santiago Escobar and
    Julia Sapi\~na}. Symbolic Analysis of Maude Theories with Narval.\\
\url{https://arxiv.org/abs/1907.10919}

\item \emph{Mario Alviano, Carmine Dodaro, Johannes K. Fichte, Markus
    Hecher, Tobias Philipp and Jakob Rath}. Inconsistency Proofs for
  ASP: The ASP-DRUPE Format.\\
\url{https://arxiv.org/abs/1907.10389}

\item \emph{Felicidad Aguado, Pedro Cabalar, Jorge Fandinno, David
    Pearce, Gilberto Perez and Concepcion Vidal}. Revisiting Explicit
  Negation in Answer Set Programming.\\
\url{https://arxiv.org/abs/1907.11467}

\item \emph{Angelos Charalambidis, Christos Nomikos and Panos
    Rondogiannis}. The Expressive Power of Higher-Order Datalog.\\
\url{https://arxiv.org/abs/1907.09820}

\item \emph{Joao Alcantara, Samy S\'a and Juan Carlos
    Acosta-Guadarrama}. On the Equivalence Between Abstract
  Dialectical Frameworks and Logic Programs.\\
\url{https://arxiv.org/abs/1907.09548}

\item \emph{Francesco Calimeri, Giovambattista Ianni, Francesco
    Pacenza, Simona Perri and Jessica Zangari}. Incremental Answer Set
  Programming with Overgrounding.\\
\url{https://arxiv.org/abs/1907.09212}

\item \emph{Thomas Eiter, Paul Ogris and Konstantin Schekotihin}. A
  Distributed Approach to LARS Stream Reasoning (System paper).\\
\url{https://arxiv.org/abs/1907.12344}

\item \emph{Jes\'us J. Doménech, John Gallagher and Samir
    Genaim}. Control-Flow Refinement by Partial Evaluation, and its
  Application to Termination and Cost Analysis.\\
\url{https://arxiv.org/abs/1907.12345}

\item \emph{Amelia Harrison and Vladimir Lifschitz}. Relating Two
  Dialects of Answer Set Programming.\\
\url{https://arxiv.org/abs/1907.12139}

\item \emph{Arpit Sharma}. Using Answer Set Programming for
  Commonsense Reasoning in the Winograd Schema Challenge.\\
\url{https://arxiv.org/abs/1907.11112}

\item \emph{Matti Berthold, Ricardo Gon\c{c}alves, Matthias Knorr and
    Joao Leite}. A Syntactic Operator for Forgetting that Satisfies
  Strong Persistence.\\
\url{https://arxiv.org/abs/1907.12501}

\item \emph{Ariyam Das and Carlo Zaniolo}. A Case for Stale
  Synchronous Distributed Model for Declarative Recursive Computation.\\
\url{https://arxiv.org/abs/1907.10278}

\item \emph{Alessio Fiorentino, Nicola Leone, Marco Manna, Simona
    Perri and Jessica Zangari}. Precomputing Datalog evaluation plans
  in large-scale scenarios.\\
\url{https://arxiv.org/abs/1907.12495}

\item \emph{Yi Wang, Shiqi Zhang and Joohyung Lee}. Bridging
  Commonsense Reasoning and Probabilistic Planning via a Probabilistic
  Action Language.\\
\url{https://arxiv.org/abs/1907.13482}

\item \emph{Joaquin Arias and Manuel Carro}. Evaluation of the
  Implementation of an Abstract Interpretation Algorithm using Tabled
  CLP.\\
\url{https://arxiv.org/abs/1908.00104}

\item \emph{David Spies, Jia-Huai You and Ryan
    Hayward}. Domain-Independent Cost-Optimal Planning in ASP.\\
\url{https://arxiv.org/abs/1908.00112}
\end{itemize}

In addition to the presentations of accepted papers, the technical
program of ICLP 2019 included four invited talks:
\begin{itemize}
\item \emph{Adnan Darwiche}. What Logic Can Do for AI Today.
\item \emph{Nicola Leone}. ASP Applications for AI and Industry.
\item \emph{Sheila McIlraith}. Reward Machines: Structuring reward
  function specifications and reducing sample complexity in
  reinforcement learning.
\item \emph{Yuliya Lierler}. System PROJECTOR: An Automatic Program
  Rewriting Tool for Non-Ground Answer Set Programs.
\end{itemize}
and three invited tutorials:
\begin{itemize}
\item \emph{Chitta Baral}. Knowledge Representation and Reasoning
  issues in Natural Language Question Answering.
\item \emph{Serdar Kadioglu}. Constraint Programming for Resource
  Management.
\item \emph{Guy Van den Broeck}. Tractable Probabilistic Circuits
\end{itemize}

Furthermore, after a thorough examination of citation indices (e.g., Web of Science, Google Scholar),
two test-of-time awards were identified:
\begin{itemize}
\item The John Alan Robinson 20 year test of time award: \emph{Fr\'ed\'eric Benhamou, Fr\'ed\'eric Goualard, Laurent Granvilliers, Jean-Francois Puget}. Revising Hull and Box Consistency. ICLP 1999: 230-244
\item The Alain Colmerauer 10 year test of time award: \emph{Martin Gebser, Max Ostrowski, Torsten Schaub}.
Constraint Answer Set Solving. ICLP 2009: 235-249
\end{itemize}

We are deeply indebted to the Program Committee members and external
reviewers, as the conference would not have been possible without
their dedicated, enthusiastic and outstanding work. The Program
Committee members of ICLP 2019 were:
\begin{center}
  \small
\begin{tabular}{llll}
  Hassan Ait-Kaci & Mario Alviano & Roman Bartak \\
  Rachel Ben-Eliyahu-Zohary & Bart Bogaerts & Gerhard Brewka \\
  Pedro Cabalar & Michael Codish & Stefania Costantini \\
  Marina De Vos & Agostino Dovier & Thomas Eiter\\
  Wolfgang Faber & Fabio Fioravanti & Andrea Formisano \\
  John Gallagher & Martin Gebser & Michael Gelfond \\
  Michael Hanus & Amelia Harrison & Manuel Hermenegildo \\
 Giovambattista Ianni & Daniela Inclezan & Katsumi Inoue\\
 Tomi Janhunen & Angelika Kimmig & Ekaterina Komendantskaya \\
 Vladimir Lifschitz & Evelina Lamma & Joohyung Lee \\
 Nicola Leone & Yanhong Annie Liu & Fred Mesnard \\
 Jose F. Morales &  Emilia Oikarinen & Carlos Olarte\\
 Magdalena Ortiz & Mauricio Osorio & Barry O'Sullivan \\
  Simona Perri & Enrico Pontelli & Ricardo Rocha \\
  Alessandra Russo & Orkunt Sabuncu & Chiaki Sakama \\
  Torsten Schaub & Guillermo R. Simari & Theresa Swift\\
 Francesca Toni & Paolo Torroni & Tran Cao Son \\
 Alicia Villanueva & Kewen Wang &  Jan Wielemaker \\
 Stefan Woltran& Fangkai Yang & Roland Yap \\
 Jia-Huai You & Zhizheng Zhang\\
\end{tabular}
\end{center}
The Program Committee members of the Applications track were:
\begin{center}
  \small
  \begin{tabular}{llll}
    Chitta Baral  & Alex Brik & Francesco Calimeri & Xiaoping Chen\\
    Federico Chesani & Mart\'{\i}n Di\'eguez & Gerhard Friedrich &
                                                                   Gopal Gupta\\
     Jianmin Ji & Gabriele Kern-Isberner & Zeynep Kiziltan & Viviana
                                                             Mascardi\\
    Yunsong Meng & Francesco Ricca & Mohan Sridharan & David Warren\\
    Shiqi Zhang & Neng-Fa Zhou\\
  \end{tabular}
\end{center}
The Program Committee members of the Special Session: Women in Logic Programming were:
\begin{center}
  \small
  \begin{tabular}{llll}
 Elvira Albert &  Stefania Costantini & Ines Dutra & Daniela Inclezan\\
    Ekaterina Komendantskaya & Simona Perri & Francesca Toni\\
  \end{tabular}
  \end{center}
The external reviewers were:
\begin{center}
  \small
  \begin{tabular}{llll}
  Van Nguyen &  Bernardo Cuteri &  Dennis Dams \\
    Anna Schuhmann & Alberto Policriti & Jessica Zangari \\
    V\'{\i}tor Santos Costa & Arash Karimi & Joxan Jaffar\\
    Michael Frank & Roland Kaminski &
                                                   Javier Romero \\
    Jose Luis Carballido & Christopher Kane &
                                                            Emanuele De Angelis \\
   Isabel Garcia-Contreras &
                             Jos\'e Abel Castellanos Joo &Wolfgang Dvorak \\
    Vitaly Lagoon & Jannik Dreier & Philipp Wanko \\
    Marco Gavanelli & Emanuel Sallinger &
                                          Weronika T. Adrian \\
    Wanwan Ren & Kinjal Basu &
                                                  Patrick Kahl \\
    Marco Alberti & Gianluca Amato &
                                     Juan Carlos Nieves \\
    Joaquin Arias & Miguel Areias & Konstantin Schekotihin \\
    Farhad Shakerin & Nada Sharaf &
                                    Christoph Redl \\
    Yuanlin Zhang & Yi Tong &
                              K. Tuncay Tekle \\
    Saksham Chand & Yan Zhang & Jessica Zangari
\end{tabular}
\end{center}

We would also like to express our gratitude to the full ICLP 2019
organization committee. Our gratitude must be extended to Torsten
Schaub, who is serving in the role of President of the Association of
Logic Programming (ALP), to all the members of the ALP Executive
Committee and to Mirek Truszczynski, Editor-in-Chief of TPLP. Also, to
the staff at Cambridge University Press for their assistance.
Finally, we wish to thank each author of every submitted paper, since
their efforts keep the conference alive and the participants to ICLP
for bringing and sharing their ideas and latest developments.

\label{lastpage}
\end{document}